\documentclass[12pt]{iopart}
\usepackage{iopams}

\input epsf.sty

\newlength{\TZ}
\newcommand{\BEQ}{\begin{equation}}     % Gleichungen Anfang ..
\newcommand{\BEA}{\begin{eqnarray}}
\newcommand{\EEQ}{\end{equation}}       % .. und Ende
\newcommand{\EEA}{\end{eqnarray}}
\newcommand{\eps}{\varepsilon}          % epsilon
\newcommand{\D}{{\rm d}}                % gerades d fuer Ableitungen
\newcommand{\wit}[1]{\widetilde{#1}}    % weite Schlange
\renewcommand{\vec}[1]{\boldsymbol{#1}} % Vektoren fettgedruckt

\newcommand{\appsection}[2]{\setcounter{equation}{0}\setcounter{subsection}{0}
\section*{Appendix #1. #2}
\renewcommand{\theequation}{#1\arabic{equation}}
              \renewcommand{\thesection}{#1} }
\catcode`\@=11
\def\numberbysection{\@addtoreset{equation}{section}
        \def\theequation{\thesection.\arabic{equation}}}
\begin{document}

\title[Scaling and ageing of the alternating 
susceptibility in spin glasses]{On the scaling and ageing behaviour 
of the alternating susceptibility in spin glasses and local scale-invariance
\footnote{Dedicated to Lothar Sch\"afer on the occasion 
of his 60$^{\rm th}$ birthday.}}

\author{Malte Henkel$^1$ and Michel Pleimling$^2$}
\address{$^1$Laboratoire de Physique des 
Mat\'eriaux,\footnote{Laboratoire associ\'e au CNRS UMR 7556} 
Universit\'e Henri Poincar\'e Nancy I, \\ 
B.P. 239, F -- 54506 Vand{\oe}uvre l\`es Nancy Cedex, France}
\address{$^2$Institut f\"ur Theoretische Physik I, 
Universit\"at Erlangen-N\"urnberg, \\
Staudtstra{\ss}e 7B3, D -- 91058 Erlangen, Germany}

\begin{abstract}
The frequency-dependent scaling of the dispersive and dissipative parts of the
alternating susceptibility is studied for spin glasses at criticality. 
An extension of the usual $\omega t$-scaling is proposed. 
Simulational data from the three-dimensional Ising spin glass agree with this
new scaling form and moreover reproduce 
well the scaling functions explicitly calculated for systems satisfying 
local scale-invariance. There is also a qualitative agreement with existing
experimental data. 
\end{abstract}

\submitto{\JPCM}
\maketitle

\setcounter{footnote}{0}

%%%%%%%%%%%%%%%%%%%%%%%%%%%%%%%%%%%%%%%%%%%%%%%%%%%%%%%%%%%%%%%%%%%%%%%%%%%%%%%%
\section{Introduction}
%%%%%%%%%%%%%%%%%%%%%%%%%%%%%%%%%%%%%%%%%%%%%%%%%%%%%%%%%%%%%%%%%%%%%%%%%%%%%%%%

Understanding the complex behaviour of glass-forming 
systems cooled to below their
glass transition temperature remains a challenge. At first sight, the problem
might appear to be hopelessly difficult, since 
time-translation invariance is in
general broken and the properties of observables may hence depend on the
prehistory of the material under study (i.e. thermal, mechanical,\ldots). 
On the other hand, an important discovery has been the observation of 
{\em dynamical scaling}, see \cite{Stru78}, which occurs quite independently 
of whether the equilibrium state is critical or not. 
In recent years, it has been realized that many aspects of dynamical scaling 
are conveniently first studied in
non-disordered, i.e. ferromagnetic systems. After a quench to or below their
critical temperature $T_c$, these systems undergo an ageing behaviour which in
many respects is quite similar to the one in glassy or kinetically constrained
systems. The manifold problems which arise in the study of ageing in
simple magnets or glasses are reviewed, 
e.g. in \cite{Bray94,Cate00,Cugl02,Godr02,Cris03,Henk04,Kawa04,Cala04}. 

For notational simplicity, we shall in what follows consider magnetic spin
systems. Convenient tools for the study of ageing behaviour are the two-time
autocorrelation and autoresponse functions
\BEA
C(t,s) &=& \langle \phi(t) \phi(s) \rangle ~~ \sim s^{-b} f_C(t/s) \\
R(t,s) &=& \left.\frac{\delta\langle\phi(t)\rangle}{\delta h(s)}\right|_{h=0}
\sim s^{-1-a} f_{R}(t/s)
\EEA
where $\phi(t)$ is the order parameter at time $t$ and $h(s)$ is the conjugate
magnetic field at time $s$. The scaling behaviour is expected to apply
in the so-called {\em ageing regime} where $t,s\gg t_{\rm micro}$ 
and $t-s\gg t_{\rm micro}$, where $t_{\rm micro}$ is a microscopic time scale.
Furthermore, we tacitly assumed that the scaling derives from the 
time-dependence of a single characteristic length-scale $L(t)$ which measures
the linear size of correlated clusters. In this paper, we shall always
consider algebraic growth laws, viz. $L(t)\sim t^{1/z}$, 
where $z$ is the dynamic exponent. Then the above forms define the 
nonequilibrium exponents $a$ and $b$ and the scaling functions $f_C(y)$ and
$f_R(y)$. For large arguments $y\to \infty$, one generically expects
\BEQ
f_C(y) \sim y^{-\lambda_C/z} \;\; , \;\;
f_R(y) \sim y^{-\lambda_R/z}
\EEQ
where $\lambda_C$ and $\lambda_R$, respectively, are known as autocorrelation
\cite{Fish88,Huse89} and autoresponse exponents \cite{Pico02}. This description
applies to many simple magnets quenched to a temperature $T\leq T_c$ as is 
well-known, see \cite{Bray94,Cate00,Godr02}.\footnote{Exceptions occur for 
example in the $2D$ XY model with a fully disordered initial state.} 
On the other hand, for glasses quenched to below the glass-transition 
temperature, a slow cross-over between an algebraic growth law at short times
towards a slower (logarithmic) growth at larger times appears to give a
better description of the data, see \cite{Kawa04} and references therein. 
Recently, evidence was found that for spin glasses quenched onto their 
critical temperature $T=T_c$ a simple scaling of the two-time
observables compatible with an algebraic growth law applies \cite{Henk04a}.

In recent years, from the study of ageing in simple magnets it has been 
proposed that global dynamical scaling might be extended
to a local scale-invariance \cite{Henk02,Henk01}. One of the 
simplest predictions of that theory is the explicit form of the two-time
autoresponse function. It reads\footnote{Technically, this requires that
the order parameter is a {\em quasi-primary} field under local 
scale-transformations \cite{Henk02}. This concept is the analogue of the one
used in conformal field-theory \cite{Scha75}.}
\BEQ \label{gl:Rf}
R(t,s) = s^{-1-a} f_R(t,s) \;\; , \;\; f_R(y) = f_0\, y^{1+a'-\lambda_R/z}
(y-1)^{-1-a'}
\EEQ
where $a'$ is a new independent exponent and $f_0$ is a normalization
constant. The independent existence of the exponent $a'$ was recognized 
recently \cite{Pico04} for $z=2$ and we shall extend that argument
to arbitrary $z$ in appendix~A. Previous derivations of $f_R(y)$ 
\cite{Henk02,Henk01} had assumed $a=a'$ from the outset. 
The only known example with
distinct exponents $a\ne a'$ was the $1D$ Glauber-Ising
model at temperature $T=0$ and with initial power-law correlations of the
form $\langle\sigma_i \sigma_j\rangle \sim |i-j|^{-\nu}$ with $\nu\geq 0$. 
The exact solution for $R(t,s)$
\cite{Godr00a,Lipp00,Henk03d} is of the form (\ref{gl:Rf}) and 
one can read off $a=0$, $a'=-\frac{1}{2}$ and $\lambda_R=1$ \cite{Pico04}. 
In appendix~B, we give a second example and show that the treatment of 
phase-ordering in $d$ spatial dimensions at zero temperature in the spirit 
of the OJK-approximation \cite{Bert99,Maze04} leads to 
\BEQ \label{gl:expo}
a_{\rm OJK} = \frac{d-1}{2} \;\; , \;\;
a_{\rm OJK}' = \frac{d-2}{2} \;\; , \;\;
\lambda_{R, {\rm OJK}} = \frac{d}{2}
\EEQ
The same scaling function and hence (\ref{gl:expo}) is also found from 
the gaussian approximation to phase-ordering kinetics 
\cite{Maze03}.\footnote{We observe that in both
examples $a=a'+\frac{1}{2}$ but it is still open to what extent this might be
a general relationship.} 

Eq.~(\ref{gl:Rf}) with $a=a'$ is perfectly reproduced in simple magnets 
undergoing coarsening after a quench to $T<T_c$, most notably the
$2D/3D$ Glauber-Ising model \cite{Henk01,Henk03b}, the $3D$ XY model
\cite{Abri04b} and in several exactly solvable 
systems \cite{Henk02,Abri04a,Pico04}.\footnote{A further extension of 
local scale-invariance with $z=2$ in $d$ spatial dimensions to a new type of 
conformal invariance in $d+2$ dimensions yields a prediction of $C(t,s)$ which
is in agreement with numerical data in the $2D$ Glauber-Ising 
model \cite{Henk04b}.} For critical quenches $T=T_c$, the
agreement between numerical data and (\ref{gl:Rf}) with $a=a'$ 
is almost perfect in the $2D/3D$ Ising and XY 
models \cite{Henk01,Abri04a,Abri04b,Plei04}, the $1D$ contact process
\cite{Enss04} and is exact in several exactly solvable 
models with $z=2$ \cite{Henk02,Pico02,Pico04}. On the other
hand, a second-order $\eps$-expansion calculation from renormalized 
field-theory gives a small but systematic deviation with respect 
to eq.~(\ref{gl:Rf}) \cite{Cala04,Plei04b}.

Of course it would be interesting to see whether a scaling description or even
an extension to local scale-invariance might be applicable to glassy systems 
as well. Such a comparison may be far from straightforward, however. Indeed,
a possible scaling behaviour of two-time correlators $C(t,s)$ 
and integrated responses $\int \!\D u\, R(t,u)$
have been discussed since a long time, both theoretically and experimentally, 
see \cite{Stru78,Cate00,Cris03,Cugl02,Dupu04,Kawa04} and debates have arisen
over several central issues of which we mention a few. 
First, for glasses quenched to below their
glass temperature, it is not entirely clear whether the growth law for $L(t)$
is algebraic or logarithmic \cite{Joen02}. Second, even if an asymptotic
power-law scaling is accepted, there has been an intense debate on whether
scaling occurs according to the so-called `full ageing' scenario, that is
in terms of the scaling variable $y=t/s$ (possibly with small logarithmic
corrections) \cite{Jime03}, or else if a `sub-ageing' scenario
applies, with a scaling variable $\xi :=[t^{1-\mu}-s^{1-\mu}]/(1-\mu)$, where
$\mu$ is a free parameter \cite{Dupu04}. 
The usual power-law scaling is recovered in the $\mu\to 1$ limit, 
but in many experiments the data are fitted with
values of $\mu$ as low as $\approx 0.8 - 0.9$. It has been 
suggested recently \cite{Rodr03} that values of $\mu<1$ merely result 
from a quench to
below the glass transition which is not yet sufficiently rapid, but the
repetition of that experiment on other substances has not yet led to 
unambiguous conclusions \cite{Dupu04}. Third, it is even no longer obvious
that the commonly studied spin glass models really mimic sufficiently well
the experimentally studied materials (in spite of well-established qualitative
similarities \cite{Picc01}): recent simulations on $3D/4D$ Ising and
Heisenberg spin glasses provide evidence for cumulative ageing and rejuvenation
phenomena in temperature cycling which are not observed in real spin glass
materials \cite{Maio04}.

In view of these many difficulties, it might be simpler to consider the 
behaviour of glassy systems from a different point of view. 
One rather works with a time-dependent (oscillating) magnetic field 
and studies simultaneously the dependence on time
and on the imposed oscillation angular frequency $\omega$. For a harmonic
magnetic field, it is common to consider the real and the imaginary part
of the magnetic susceptibility
\BEA
\chi'(\omega,t) = \int_0^t \!\D u\: R(t,u) \cos\left( \omega(t-u)\right)
\nonumber \\
\chi''(\omega,t) = \int_0^t \!\D u\: R(t,u) \sin\left( \omega(t-u)\right)
\label{gl:sus}
\EEA
where $R(t,s)$ is the linear response discussed above. In this setting, 
$1/\omega$ provides the second time-scale, the natural scaling variable
is $y=\omega t$ and the scaling regime should be reached in the
limit $\omega\to 0$ and $t\to\infty$. In many experiments and simulations,
one averages over at least one period of the oscillating field, 
see e.g. \cite{Picc01}. Then in a great variety
of glass-forming substances quenched to below or near to 
their glass transition point
one observes good but not always perfect evidence for an
$\omega t$-scaling behaviour  
of the following form of the period-averaged dissipative
(imaginary) part \cite{Joen02,Suzu03}
\BEQ \label{gl:chi2p}
\overline{\chi''}(\omega,t) = 
\chi''_{\rm st}(\omega) + \chi''_{\rm age}(\omega,t)
\;\; , \;\; 
\chi''_{\rm age}(\omega,t) \simeq A''_{\rm age} \left( \omega t\right)^{-b''}
\EEQ
where $\chi''_{\rm st}$ is thought of as a `stationary' contribution while
the ageing behaviour is described by $\chi''_{\rm age}$. The amplitude
$A''_{\rm age}$ and the exponent $b''$ are obtained from fits to the
experimental data but there does not seem to exist a relationship with the
exponents $a,a',b,\lambda_{C,R}$ in the literature. 

Similar scaling forms have been proposed for the dispersive (real) 
part $\overline{\chi'}$ but in practice the imaginary part is usually easier to 
measure. It is usually thought that $b'=b''$. 

In this paper, we shall try and see whether a relationship between the
exponents $b''$ and $b'$ of the alternating susceptibilities (\ref{gl:sus})
and the other nonequilibrium exponents arising in the two-time observables 
can be found. Assuming the validity of local scale-invariance for the
relatively large values of $z$ (usually, $z\approx 5-7$ \cite{Kawa04}) 
found in many studies
of spin glasses and hence the explicit form (\ref{gl:Rf}) for the autoresponse
function,\footnote{Our recent study of the critical $3D/4D$ Ising spin glass
\cite{Henk04a} showed that the form of the measured thermoremanent 
magnetization $\rho(t,s)=\int_0^s\!\D u\, R(t,u)$ agrees with the prediction 
of local scale-invariance for $t/s\lesssim 20$, which is also in the sector 
encountered in $\omega t$-scaling.} we shall show that  for large times
\BEA
\chi''(\omega,t) = \chi''_1(\omega) + \omega^{a} \chi''_2(\omega t) + 
{\rm O}\left( t^{-\lambda_R/z}\right) 
\nonumber \\
\chi'(\omega,t) = \chi'_1(\omega) + \omega^{a} \chi'_2(\omega t) + 
{\rm O}\left( t^{-\lambda_R/z}\right) 
\label{chiSkal}
\EEA
and we shall calculate the scaling functions $\chi''_2$ and $\chi'_2$
explicitly in section~2. For the asymptotics of these scaling functions
we expect $\chi'_2(y)\sim y^{-b'}$ and $\chi''_2(y)\sim y^{-b''}$ 
for $y\gg 1$ and obtain the relation
\BEQ \label{gl:bbaa}
b' = b'' = a - a'
\EEQ
This should be compared to the experimentally found scaling (\ref{gl:chi2p}) of
$\chi''_{\rm age}(\omega t)$ and similarly for $\chi'_{\rm age}(\omega t)$. 
In section~3 we compare these results with Monte-Carlo data from the $3D$ Ising
spin glass with a binary distribution of the couplings
and discuss to what extent the scaling relations (\ref{gl:bbaa})
and the explicit scaling functions agree with existing experimental data. 
We conclude in section~4. In appendix~A we derive eq.~(\ref{gl:Rf}) from
(an extension of) local scale-invariance and in appendix~B we revisit the
OJK-approximation of coarsening kinetics and derive (\ref{gl:expo}).

%%%%%%%%%%%%%%%%%%%%%%%%%%%%%%%%%%%%%%%%%%%%%%%%%%%%%%%%%%%%%%%%%%%%%%%%%%%%%%%%
\section{Scaling of the alternating susceptibility}
%%%%%%%%%%%%%%%%%%%%%%%%%%%%%%%%%%%%%%%%%%%%%%%%%%%%%%%%%%%%%%%%%%%%%%%%%%%%%%%%

We now analyze the frequency-dependent scaling of the alternating 
susceptibility. For notational simplicity, we concentrate first on 
$\chi''(\omega,t)$ as given by eq.~(\ref{gl:sus}). In order to make the
scaling behaviour explicit, we must convert this into a more convenient
form which can be done as follows \cite{Henk02a}. We observe that the time
difference $\tau=t-u$ plays a central r\^ole since depending on its value
either an equilibrium behaviour or else an ageing behaviour is obtained.
Specifically, it can be shown \cite{Zipp00} that there is a time-scale
$t_p\sim t^{\zeta}$ with $0<\zeta<1$ on which the transition between the
two regimes occurs such that $R(t,s)\simeq R_{\rm eq}(t-s)$ for 
$t-s\lesssim t_p$ and $R(t,s)=s^{-1-a}f_R(t/s)$ as given in eq.~(\ref{gl:Rf}) 
for $t-s\gtrsim t_p$.\footnote{Explicitly, $\zeta=4/(d+2)$ in the 
$d$-dimensional spherical model \cite{Zipp00}.} 
In addition, one measures for $u\approx t$ the response
with respect to a change in the initial conditions and then instead of
(\ref{gl:Rf}) one expects $R\approx R_{\rm ini}(t)\sim t^{-\lambda_R/z}$ 
\cite{Bray94}. We therefore must introduce a further time-scale $t_{\eps}$ 
such that $t-t_{\eps}={\rm O}(1)$. Changing variables and then splitting the
integral into three terms corresponding to these three regimes, we have
\BEA
\lefteqn{ \chi''(\omega,t) = \int_0^t \!\D\tau\: R(t,t-\tau) \sin \omega\tau}
\nonumber \\
&=& \int_0^{t_p} \!\D\tau\: R(t,t-\tau) \sin \omega\tau +
\int_{t_p}^{t_{\eps}} \!\D\tau\: R(t,t-\tau) \sin \omega\tau 
\nonumber \\
& & +\int_{t_{\eps}}^t \!\D\tau\: R(t,t-\tau) \sin \omega\tau 
\nonumber \\
&\simeq& \int_{0}^{t_p} \!\D\tau\: R_{\rm eq}(\tau) \sin\omega\tau +
t^{-a}\int_{t_p/t}^{t_{\eps}/t} \!\D v\: f_R\left(\frac{1}{1-v}\right)
\frac{\sin \omega tv}{(1-v)^{1+a}}
\nonumber \\ 
& & +t^{-\lambda_R/z} \int_{t_{\eps}}^{t} \!\D\tau\: c_0 \sin\omega\tau
\nonumber \\
&=& \chi_1''(\omega) +
t^{-a}\int_{0}^{1} \!\D v\: f_R\left(\frac{1}{1-v}\right)
\frac{\sin \omega tv}{(1-v)^{1+a}} +
{\rm O}\left(t^{-\lambda_R/z}\right) \label{chipp}
\EEA
In the third line, we used the asymptotic forms of $R(t,s)$ as described
above. This means that the cross-over between the equilibrium and the ageing 
regimes is assumed to be very rapid. In the last line, we restricted 
ourselves to the long-time
limit $t\to\infty$. Here, the function $\chi_1''(\omega)$ merely depends on the
equilibrium form of the linear response $R_{\rm eq}(t,s)$. 

In this way (analogously for $\chi'$) the scaling form (\ref{chiSkal}) is
obtained. This derivation also shows that the often-found stationary term 
in the integrated response 
\cite{Cate00,Coll00,Cugl02,Dupu04,Joen02,Kawa04,Rhei04,Rodr03,Suzu03} 
does not require the separation of a similar `stationary' part in the response
function $R(t,s)$ itself.

We now analyze the second term in the above expression for $\chi''$. Using
the explicit form (\ref{gl:Rf}) for the scaling function $f_R$, we have
\BEQ
\chi''(\omega,t) = \chi_1''(\omega) + f_0 t^{-a} S 
+ {\rm O}\left(t^{-\lambda_R/z}\right)
\EEQ
where expansion of the sine followed by term-wise integration gives
\BEA
\lefteqn{ \hspace{-2truecm} 
S := \int_0^1 \!\D v\: (1-v)^{-1-a+\lambda_R/z} v^{-1-a'}
\sin (\omega t v)}
\nonumber \\
&\hspace{-2truecm}=& \hspace{-1.5truecm} 
\sum_{n=0}^{\infty} \frac{(-1)^n}{(2n+1)!} 
B\left(2n+1-a',\frac{\lambda_R}{z}-a\right) \,\left( \omega t\right)^{2n+1}
\nonumber \\
&\hspace{-2truecm}=& \hspace{-1.5truecm} 
B\left( 1-a', \frac{\lambda_R}{z}-a\right) \,\omega t
\\
& & \hspace{-1.5truecm} \times 
{_2F_3}\left(\frac{1-a'}{2},\frac{2-a'}{2};\frac{3}{2},
\frac{1-a-a'}{2}+\frac{\lambda_R}{2z},\frac{2-a-a'}{2}+\frac{\lambda_R}{2z};
-\frac{\omega^2 t^2}{4}\right) 
\nonumber 
\EEA
where ${}_2F_3$ is a hypergeometric function,  
\BEQ
B(z,w) = \frac{\Gamma(z)\Gamma(w)}{\Gamma(z+w)} 
=\int_0^1 \!\D u\: u^{z-1} (1-u)^{w-1}
\EEQ
is Euler's betafunction and the identity 
$\Gamma(2z)=2^{2z-1}\Gamma(z)\Gamma(z+1/2)/\sqrt{\pi}$ was also used. 

We proceed
to extract the leading behaviour for large values of the scaling variable
$y=\omega t$. Recall first the asymptotic expansion 
for $x\to+\infty$ \cite{Wrig35}
\BEA
\hspace{-2truecm}{_2F_3}(a,b;c,d,e;-x) &\simeq& \left[ 
A x^{-a} + B x^{-b} + C x^{\Delta/2} 
\cos\left( 2\sqrt{x\,}+\frac{\pi}{2}\Delta\right)
\right] 
\nonumber \\
&\times&\left( 1 + \mbox{\rm O}\left( x^{-1}\right)\right)
\EEA
with $\Delta=a+b-c-d-e+\frac{1}{2}$ and where the constants 
$A,B,C$ are given by
\BEA
A &=& \frac{\Gamma(c)\Gamma(d)\Gamma(e)}{\Gamma(b)}
\frac{\Gamma(b-a)}{\Gamma(c-a)\Gamma(d-a)\Gamma(e-a)} 
\nonumber \\
B &=& \frac{\Gamma(c)\Gamma(d)\Gamma(e)}{\Gamma(a)}
\frac{\Gamma(a-b)}{\Gamma(c-b)\Gamma(d-b)\Gamma(e-b)} 
\nonumber \\
C &=& \frac{\Gamma(c)\Gamma(d)\Gamma(e)}{\sqrt{\pi\,}\Gamma(a)\Gamma(b)}
\EEA
Inserting this into the expression for $S$, we find
\BEQ \label{gl:S}
\hspace{-0.5truecm} S \simeq \bar{A} \left( \omega t\right)^{a'-a} + 
\bar{B} \left( \omega t\right)^{a'-a-1} +
\bar{C} \left(\omega t\right)^{-\lambda_R/z} \sin\left(\omega t+ 
\frac{\pi}{2}\left(a-\lambda_R/z\right)\right)
\EEQ
where the constants $\bar{A},\bar{B},\bar{C}$ are proportional to $A,B,C$. 
The second term in (\ref{gl:S}) is always non-leading. The other two terms,
however, will determine the functional form of the scaling function for
$y=\omega t$ sufficiently large. The treatment of $\chi'$ is analogous. 

We can summarize the content of this section by listing the scaling functions
which occur in (\ref{chiSkal}), together with their leading behaviour 
as $y\to\infty$
\BEA
\hspace{-2.0truecm}\chi_2''(y) &\hspace{-0.9truecm}=& \hspace{-0.3truecm}
f_0 B\left(1-a',\frac{\lambda_R}{z}-a\right) \,y^{1-a} 
\nonumber \\
& & \hspace{-1.5truecm} \times
{_2F_3}\left(\frac{1-a'}{2},\frac{2-a'}{2};\frac{3}{2},
\frac{1-a-a'}{2}+\frac{\lambda_R}{2z},\frac{2-a-a'}{2}+\frac{\lambda_R}{2z};
-\frac{y^2}{4}\right) 
\label{3:gl:cc} \\
&\hspace{-1.8truecm}\simeq& \hspace{-1.3truecm} 
f_0 \frac{\pi}{2}\left[\cos\left(\frac{\pi a'}{2}\right)\Gamma(1+a')
\right]^{-1} y^{a'-a} + f_0 \Gamma\left(\frac{\lambda_R}{z}-a\right)
y^{-\lambda_R/z}\sin\left(y+\frac{\pi}{2}\left[a-\lambda_R/z\right]\right)
\nonumber \\
\hspace{-2.0truecm}\chi_2'(y) &\hspace{-0.9truecm}=&  \hspace{-0.3truecm} 
f_0 B\left(-a',\frac{\lambda_R}{z}-a\right) \,y^{-a}
\nonumber \\
& & \hspace{-1.5truecm} \times
{_2F_3}\left(\frac{-a'}{2},\frac{1-a'}{2};\frac{1}{2},
\frac{-a-a'}{2}+\frac{\lambda_R}{2z},\frac{1-a-a'}{2}+\frac{\lambda_R}{2z};
-\frac{y^2}{4}\right)
\label{3:gl:ccc}\\
&\hspace{-1.8truecm}\simeq& \hspace{-1.3truecm} 
-f_0 \frac{\pi}{2} \left[\sin\left(\frac{\pi a'}{2}\right)\Gamma(1+a')
\right]^{-1} y^{a'-a} + f_0 \Gamma\left(\frac{\lambda_R}{z}-a\right)
y^{-\lambda_R/z}\cos\left(y+\frac{\pi}{2}\left[a-\lambda_R/z\right]\right)
\nonumber 
\EEA

We see that there appear terms which decrease monotonously with $y$ 
but that there are
also oscillating terms. They are described by different exponents and must
be extracted by a different experimental setup. The oscillating terms follow
the oscillations of the external field and the decrease of the oscillation
amplitude gives a direct access to the exponent $\lambda_R/z$. 
On the other hand, in many experiments the data are averaged over one or several
periods of the external field. For $y$ sufficiently large, the contribution of
the oscillating term in eqs.~(\ref{3:gl:cc},\ref{3:gl:ccc}) vanishes after
averaging and then only
a simple algebraic component remains, which permits to extract the exponent
$a-a'$. For period-averaged data or else if $\lambda_R/z\geq a-a'$, the
leading behaviour for large arguments is
\BEQ
\chi_2'(y) \sim \chi_2''(y) \sim y^{a'-a}
\EEQ
and the scaling relations (\ref{gl:bbaa}) follow. 

%%%%%%%%%%%%%%%%%%%%%%%%%%%%%%%%%%%%%%%%%%%%%%%%%%%%%%%%%%%%%%%%%%%%%%%%%%%%%%%%
\section{Tests}
%%%%%%%%%%%%%%%%%%%%%%%%%%%%%%%%%%%%%%%%%%%%%%%%%%%%%%%%%%%%%%%%%%%%%%%%%%%%%%%%

\subsection{Numerical simulations} 

We now compare the theory of the alternating susceptibility developed in the
previous section with numerical simulations performed on the critical 
three-dimensional Ising spin glass. The Hamiltonian of the model is given by
\BEQ
{\cal H} = - \sum_{(\vec{i},\vec{j})} 
J_{\vec{i},\vec{j}} \sigma_{\vec{i}} \sigma_{\vec{j}}
\EEQ
where the nearest-neighbour couplings $J_{\vec{i},\vec{j}}$ are random 
variables taken from a binary distribution, i.e.\ $J_{\vec{i},\vec{j}}= \pm 1$ 
with equal probability. The classical Ising spins $\sigma_{\vec{i}}=\pm 1$ 
characterize the local magnetization at the sites $\vec{i}$ of a
simple cubic lattice. In the past,
numerical studies \cite{Mari02,Ball00} investigated the static critical
properties of this model which undergoes a continuous phase transition at the
temperature $T_c \approx 1.19$ (setting $k_B=1$). Recently, we studied
the ageing behaviour of this critical spin glass \cite{Henk04a}. There we
found clear evidence of a power-law scaling in the two-time correlation
and integrated response functions. Here we shall
need the exponent estimates $a = 0.060(4)$ and $\lambda_R/z = 0.38(2)$ obtained
from a scaling analysis of the thermoremanent magnetization. 

In order to study the alternating susceptibility far from equilibrium
we prepared the system in an uncorrelated initial state (corresponding
to an infinite initial temperature) before quenching it to the 
critical point at time $t=0$. At the same time an external oscillating
magnetic field 
\BEQ \label{gl:h}
h(t) = h_0 \cos \omega t
\EEQ
was switched on, with it's amplitude fixed at $h_0=0.05$ which is well
inside the linear-response regime. We consider
different values of the angular frequency $\omega = 2 \pi/p$
with $p$ ranging from $50$ to $1600$. Typically, systems containing
$50^3$ spins were simulated.

Numerically, the in-phase and the out-of-phase susceptibilities
are given by the expressions \cite{Ande96}
\BEA
\chi''(\omega,t) & =  & m(t) \sin \omega t \nonumber \nonumber \\
\chi'(\omega,t) & =  & m(t) \cos \omega t \nonumber 
\EEA
with $m(t) = \sum\limits_{\vec{i}} \sigma_{\vec{i}}(t)$.
In order to access the scaling parts of 
$\chi''$ and of $\chi'$, we must first subtract the equilibrium parts
$\chi''_{1}$ and $\chi'_{1}$. We therefore carried out longer runs
where we let the system equilibrate for typically a few ten thousand
time steps before switching on the oscillating field. When no change
in the amplitudes of the alternating susceptibilities were observed,
we identified these data with the equilibrium parts $\chi''_{1}$ and 
$\chi'_{1}$. The ageing parts of $\chi''$ and of $\chi'$
discussed in the following result from averaging over 2500 
different runs with different couplings, different initial states and 
different realizations of the thermal noise.

%%----------------------------------------------------------------------------%%
\begin{figure}
\centerline{\epsfxsize=4.0in\epsfbox
{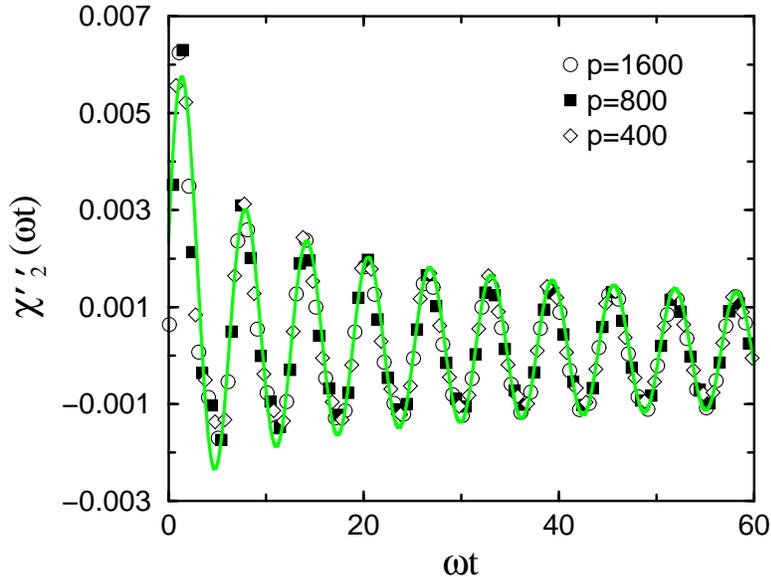}}
\caption{Scaling of the dissipative part $\chi_2''(\omega t)$ 
of the alternating susceptibility as function of the 
scaling variable $\omega t$ for different angular
frequencies $\omega= 2 \pi/p$ with $p= 1600$, 800, and 400. The full curve
is the theoretical prediction (\ref{3:gl:cc}) with 
$f_0 = 0.002$ and $a' = -0.70$ but which has also been shifted 
horizontally by $y\to y+\Delta y$, with $\Delta y= -0.45$, see text. 
Statistical error bars are smaller than the symbol sizes.
\label{fig1}}
\end{figure}
%%----------------------------------------------------------------------------%%

%%----------------------------------------------------------------------------%%
\begin{figure}
\centerline{\epsfxsize=4.0in\epsfbox
{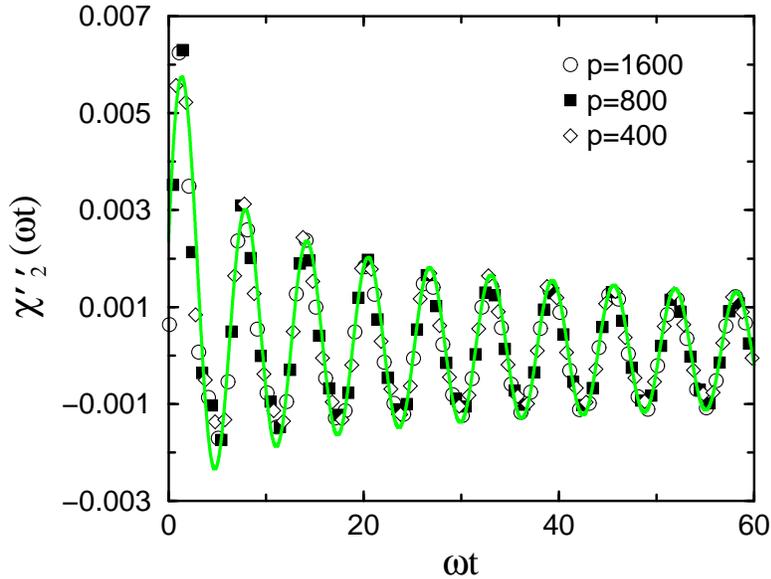}}
\caption{The same as in Figure \ref{fig1}, but now for the dispersive part 
$\chi_2'(\omega t)$. The full curve is the shifted theoretical prediction 
(\ref{3:gl:ccc}) with $f_0 = 0.002$ and $a' = -0.70$.
\label{fig2}}
\end{figure}
%%----------------------------------------------------------------------------%%

In figures \ref{fig1} and \ref{fig2} we test the expected
scaling behaviour of the ageing part, see eq.~(\ref{gl:chi2p})
\BEQ
\chi_2'' = \chi_2''(\omega t) ~~~ \mbox{and} ~~~
\chi_2' = \chi_2'(\omega t).
\EEQ
For the larger values of $p$, corresponding to the smaller values
of $\omega$, we observe a very good data collapse for both quantities, which
furnishes clear evidence in favour of a power-law scaling at $T=T_c$. 
Here we used the value $a=0.060(4)$ determined
previously from the decay of the
thermoremanent magnetization \cite{Henk04a}. 
For smaller values of $p$, the collapse is less good, which presumably means 
that for the corresponding values
of $\omega$ the dynamical scaling regime is not yet reached.

The data shown in these two figures can in principle be compared directly
with the analytical predictions (\ref{3:gl:cc}) and (\ref{3:gl:ccc}). 
We point out, however, that the positions 
of maxima of $\chi_2''$ and
$\chi_2'$ are shifted by the amount $\Delta y\approx -0.45$ when compared with
the positions obtained in the analytical treatment of the previous section.
The origin of this small shift is not completely clear to us. It is
possible, however, that the cross-over
between the equilibrium and the ageing regimes is not almost instantaneous
in contrast to what we assumed in the derivation of eq.~(\ref{chipp}). 
In any case, we shifted
the scaling variable in the analytical curves correspondingly, in 
order to make a comparison between analytical prediction and 
numerical data possible.

As both $a$ and $\lambda_R/z$
are known from our earlier investigation \cite{Henk04a} the only
free parameters in this comparison are the amplitude $f_0$
and the exponent $a'$. 
Our final estimates are
\BEQ \label{3:gl:af}
a' = -0.70(3) \;\; , \;\; f_0 = 0.00203(1) 
\EEQ
and the fit is compared to the data in the figures. 
These values (\ref{3:gl:af}) of the parameters describe consistently 
both $\chi_2''$ and $\chi_2'$.
Although some discrepancies are observed for small values of $y=\omega t$,
the overall agreement between the simulation and the shifted theoretical
prediction is very good. The numerical data therefore
support the scaling approach presented in the previous section.

\subsection{Comparison with experiments} 

We now turn to a discussion of existing experimental results on the scaling
of the alternating susceptibility. 

A detailed discussion of a possible scaling of the alternating susceptibility
was presented by Suzuki and Suzuki for the short-ranged Ising spin glass
Cu$_{0.5}$Co$_{0.5}$Cl$_2$-FeCl$_3$ graphite bi-intercalation 
compound \cite{Suzu03}. After a rapid quench to below the glass temperature
$T_g=3.92(11) {\rm K}$, they measure $\chi''(\omega,t)$ for fixed $\omega$ and
find that their (period-averaged) data are well-fitted by the power-law
\BEQ
\overline{\chi''}(\omega,t) = \chi_0''(\omega) + A''(\omega) t^{-b''}
\EEQ
While the fitted exponent $b''$ depends only slightly on $\omega$, they further
show evidence for a power-law $A''(\omega)=A_0'' \omega^{-\mu''}$ and find
that {\it ``\ldots the value of $\mu''$ is almost the same as that of 
$b''$''} \cite{Suzu03}.
In this way, they arrive at the $\omega t$-scaling form
\BEQ
\overline{\chi''}(\omega,t) = 
\chi_0''(\omega) + A_0'' \left( \omega t\right)^{-b''}
\EEQ
and a similar form for $\overline{\chi'}(\omega,t)$ 
where {\it ``\ldots $b'$ and $b''$ 
are of the same order at the same temperature''} \cite{Suzu03}.
Experimentally measured values of the exponent $b''$ (and also $b'$) of some
materials are collected in table~\ref{table1}. We now compare the
experimental results of Suzuki and Suzuki \cite{Suzu03} with the theoretical
scaling form (\ref{chiSkal}). First, the experimental evidence for a pure
$\omega t$-scaling indicates that the exponent $a$ must indeed be very small. 
Second, when considering the leading behaviour for $y=\omega t$ large 
(their data go up to $y\lesssim 10^6$ \cite{Suzu03}) and recalling that 
the experimental data are averaged over at least one period of the 
external field, we can read off  
\BEQ
b''= a - a' \;\; , \;\; \mu'' = -a'
\EEQ
and the observed \cite{Suzu03} near equality $b''\approx \mu''$ is 
again consistent with $a$ being negligibly small. Third, the available data
are consistent with the theoretically requested relation $b'=b''$. Forth, the
reason for the experimentally observed sudden jump 
in $b'$ and $b''$ for smaller $T$ is not yet understood but we
mention that a similar phenomenon also occurs in certain relaxor 
ferroelectrics \cite{Coll00,Coll01}. 

%%++++++++++++++++++++++++++++++++++++++++++++++++++++++++++++++++++++++++++++++
\begin{table}
\caption{Measured values of the exponents $b''$ and $b'$ in several glassy
materials, using the scaling form (\ref{gl:chi2p}). 
Here $T_g$ stands for the glass transition temperature and 
$T$ is the temperature where the data were taken. 
For Fe$_{0.5}$Mn$_{0.5}$TiO$_3$ and CdCr$_{1.7}$In$_{0.3}$S$_4$ the
relation $b'=b''$ was assumed.\\ \label{table1}}
%\begin{center}
\hspace{-1truecm}\begin{tabular}{|ll|llll|ll|} \hline
\multicolumn{2}{|c|}{Material} & \multicolumn{1}{c}{$T_g \mbox{\rm [K]}$} & 
\multicolumn{1}{c}{$T \mbox{\rm [K]}$} & \multicolumn{1}{c}{$b''$} & 
\multicolumn{1}{c|}{$b'$} & Ref. & \\ \hline
\multicolumn{2}{|l|}{Cu$_{0.5}$Co$_{0.5}$Cl$_2$-FeCl$_3$} & 
3.92(11) & 3.25 & 0.01(4) & 0.08(3) & \cite{Suzu03} & Ising spin glass\\
\multicolumn{2}{|l|}{-- GBIC} &     & 3.5  & 0.017(32) & 0.05(2) & & \\
 & &     & 3.75 & 0.16(3)   & 0.20(2) & & \\
 & &     & 3.85 & 0.15(3)   &         & & \\
 & &     & 3.95 & 0.16(4)   & 0.20(2) & & \\ \hline 
\multicolumn{2}{|l|}{Fe$_{0.5}$Mn$_{0.5}$TiO$_3$} & 
20.7     & 15   & 0.14(3)   & & \cite{Dupu01,Dupu02} & Ising spin glass \\
 & &     & 19   & 0.14(3)   &         & & \\ \hline
\multicolumn{2}{|l|}{CdCr$_{1.7}$In$_{0.3}$S$_4$} & 
16.7     & 12   & 0.18(3)   & & \cite{Dupu01,Dupu02} & Heisenberg spin glass \\
 & &     & 14   & 0.18(3)   &         & & \\ \hline 
CdCr$_{2x}$In$_{2-2x}$S$_4$ & $x=0.95$ & 
70       & 8    & 0.2       & & \cite{Dupu02a,Dupu02} & disordered ferromagnet\\
 & &     & 67   & 0.2       &         & & \\
 & $x=0.90$ 
         & 50 & 42 & 0.20   &         & \cite{Vinc00,Dupu02} & \\ \hline 
\multicolumn{2}{|l|}{Pb(Mg$_{1/3}$Nb$_{2/3}$)O$_3$} & 
$\sim 220$ & $\lesssim 220$ & 0.17 &  & \cite{Coll00} & relaxor ferroelectric 
\\ \hline 
\end{tabular}%\end{center}
\end{table}
%%++++++++++++++++++++++++++++++++++++++++++++++++++++++++++++++++++++++++++++++

Similar values of $b''$ were observed for several other materials, quite
independently of the precise physical nature as can be seen from 
table~\ref{table1}, but the errors are still too large to permit a discussion
of the universality of the exponents. 
However, the experimental data are in many of these
systems at least as well described by a logarithmic scaling as expected from
the droplet theory \cite{Dupu02,Joen02,Dupu01}. Furthermore, in several
systems also strong deviations from a simple $\omega t$-scaling were found,
see \cite{Coll01}. Finally, we mention that in systems like 
$\beta$-hydroquinol-clathrate \cite{Rhei04} or even simple liquids like
glycerol \cite{Lehe98} a power-law dependence of the form 
$\chi''_{\rm age}\sim t^{-a}$ or $\chi'_{\rm age}\sim t^{-a}$ was observed. 
All in all, it is not yet completely understood what precise conditions are
needed such that a clear power-law scaling in $\omega t$ can be observed. 

Lastly, we see that the value $b''\simeq 0.7$ obtained from the critical
$3D$ Ising spin glass with binary disorder is very far from the values
$b''\approx 0.1 - 0.2$ found experimentally.

%%%%%%%%%%%%%%%%%%%%%%%%%%%%%%%%%%%%%%%%%%%%%%%%%%%%%%%%%%%%%%%%%%%%%%%%%%%%%%%%
\section{Conclusions and discussion}
%%%%%%%%%%%%%%%%%%%%%%%%%%%%%%%%%%%%%%%%%%%%%%%%%%%%%%%%%%%%%%%%%%%%%%%%%%%%%%%%

The objective of this work has been to investigate to what extent the
ageing behaviour of the dispersive and dissipative parts of the 
alternating susceptibility in glassy systems may be described in terms of
some simple ideas borrowed from the ageing of simple magnets without disorder.
These are
\begin{enumerate}
\item the clear separation of the stationary and the ageing regimes, which 
goes into the derivation of eq.~(\ref{chipp}).
\item the hypothesis of a single essential length scale $L(t)\sim t^{1/z}$ 
growing algebraically in time. 
\item the extension of this dynamical scaling to a local scale-invariance 
which leads to the simple form eq.~(\ref{gl:Rf}) for the two-time response
function $R(t,s)$. 
\end{enumerate}
Taking these assumptions as working hypothesis, our results are as follows:
\begin{enumerate}
\item The often-studied simple $\omega t$-scaling (\ref{gl:chi2p}) 
in $\chi''$ and $\chi'$
should be slightly generalized to the scaling forms (\ref{chiSkal}) or
equivalently 
\BEQ \label{4:gl:ccc}
\chi''(\omega,t) = \chi''_1(\omega) + t^{-a} \bar{\chi}''_2(\omega t) \;\; 
\mbox{\rm ~with~} \;\; \bar{\chi}''_2(y)= y^{a} {\chi}''_2(y)
\EEQ
and similarly for $\chi'$. The smallness of the exponent $a$ makes it 
conceivable that the slight `subageing' found in many experiments might be
taken into account this way.\footnote{Working with (\ref{4:gl:ccc}) rather than
with (\ref{chiSkal}) has the advantage that for $y\gg 1$ the first term
in the asymptotic expansion only depends on the exponent $a'$, while the
second one only depends on $a-\lambda_R/z$.} Tests of this idea in real
materials would be welcome. 
\item The exponents $b'$ and $b''$ defined from $\chi_2'(y)\sim y^{-b'}$ and
$\chi_2''(y)\sim y^{-b''}$ satisfy the scaling relation
\BEQ 
b'=b'' = a-a'
\EEQ
which hence relates exponents found in an oscillating field to those 
describing the two-time response $R(t,s)$. Since $a'$ appears to be a new, 
independent exponent, this suggests that the exponent $b'=b''$ should 
be independent of the exponents
$a,b,\lambda_C/z,\lambda_R/z$ habitually used in describing ageing. 
This question could be addressed, independently of the hypothesis of local
scale-invariance and the form (\ref{gl:Rf}), 
through field-theory methods as developped in \cite{Pime02}.  
\item When testing the explicit dispersive and dissipative 
scaling functions (\ref{3:gl:cc}) and (\ref{3:gl:ccc})
on the $3D$ critical Ising spin glass with binary disorder, we found that
the form of {\em both} is reproduced by the data for the following exponents
\BEQ
a = 0.060(4) \;\; , \;\; 
a' = -0.70(3) \;\; , \;\;
\frac{\lambda_R}{z} = 0.38(2)
\EEQ
(and with a normalization constant $f_0\simeq 0.002$) 
where we took over the values of $a$ and $\lambda_R/z$ from our earlier 
analysis of $R(t,s)$ in this model \cite{Henk04a}. We stress that this 
agreement is {\em only} found if the 
curves (\ref{3:gl:cc}) and (\ref{3:gl:ccc})
are shifted $y\to y+\Delta y$ by a constant $\Delta y\simeq -0.45$. Even then,
for small values of $y$ the simulated scaling functions show a small but
systematic deviation from the theoretical prediction which remains to be
understood. 

The origin of this shift is unknown to us, but it might be due to a
rather slow cross-over between the stationary and the ageing regimes which is
not captured by eq.~(\ref{chipp}). Understanding this point is an
important open problem. 

We think it is remarkable that the ageing dynamics of a system as complicated
as a critical spin glass can be captured by our relatively simple hypothesis
of a dynamical symmetry. 
\item The value $b''\simeq 0.7$ we found in the $3D$ Ising spin glass 
(with binary disorder) is far
from the values obtained in experiments, 
see table~\ref{table1}.\footnote{Our previous investigation of 
the thermoremanent magnetization \cite{Henk04a} did not permit
us to determine the value of $a'$.}
If the present result
should be confirmed, it might furnish a further hint towards a fundamental
difference between simple spin glass models and experimentally realized glassy
systems \cite{Maio04}. 

On the other hand, the theoretically predicted relation $b'=b''$ is fully
consistent with the available experimental results. 
\end{enumerate}

{\bf Acknowledgements:} We thank G.F. Mazenko for helpful correspondence and
O. Petracic for a useful discussion. 
This work was supported by the Bayerisch-Franz\"osisches Hochschulzentrum (BFHZ)
and by CINES Montpellier (projet pmn2095). MP acknowledges the support by
the Deutsche Forschungsgemeinschaft through grant no. PL 323/2.

%\newpage

%%%%%%%%%%%%%%%%%%%%%%%%%%%%%%%%%%%%%%%%%%%%%%%%%%%%%%%%%%%%%%%%%%%%%%%%%%%%%%%%
\appsection{A}{The autoresponse function and extended local scale-invariance}
%%%%%%%%%%%%%%%%%%%%%%%%%%%%%%%%%%%%%%%%%%%%%%%%%%%%%%%%%%%%%%%%%%%%%%%%%%%%%%%%

Local scale-transformations are extensions of the dynamical 
scale-transformation $\vec{r}\mapsto b\vec{r}$, $t\mapsto b^z t$
of space-time towards variable
rescaling factors $b=b(t,\vec{r})$ such that conformal transformation in
time $t\mapsto (\alpha t+\beta)/(\gamma t+\delta)$ with 
$\alpha\delta-\beta\gamma=1$ are maintained \cite{Henk02}. For any given 
value of $z$, such infinitesimal transformations have been explicitly 
constructed and shown to furnish a dynamical symmetry of the generalized
diffusion equation $\left(\partial_t +\partial_{r}^{z}\right)\psi(t,r)=0$. 
Here we are interested in applications to ageing behaviour of the autoresponse
function
\BEQ
R(t,s) = \left. \frac{\delta\langle\phi(t,\vec{r})\rangle}{\delta h(s,\vec{r})}
\right|_{h=0} = \left\langle \phi(t,\vec{r})\wit{\phi}(s,\vec{r})\right\rangle
\EEQ
where $\phi$ is the order parameter, $h$ the conjugate magnetic field and
$\wit{\phi}$ the associate response field in the context of the 
Martin-Siggia-Rose theory, see e.g. \cite{Jans92}. 
Therefore, we have to restrict to the subalgebra
with time-translations excluded \cite{Henk02,Henk01}. 
Since space translations are included in the set of local 
scale-transformations, the scaling of $R(t,s)$ will be entirely given in terms
of merely two generators which we write as
\BEQ
X_0 = - t\partial_t - \frac{x}{z}
\;\; , \;\; 
X_1 = -t^2\partial_t -\frac{2}{z}\left( x+\xi\right)t
\EEQ
where $x$ is the scaling dimensions of the field on which the generators
$X_{n}$ act. We have observed earlier \cite{Pico04} in the special case $z=2$
that a further constant $\xi$ can be introduced without changing the commutator
relations of the subalgebra of local-scale invariance under consideration and
now extend this to arbitrary values of $z$. Previous treatments of the 
question had admitted $\xi=\wit{\xi}=0$ from the outset. 
The covariance of $R(t,s)$ is
now expressed as usual \cite{Henk02,Henk01}
\BEA
X_0 R &=& \left( -t\partial_t -s\partial_s -\frac{x}{z}-\frac{\wit{x}}{z}
\right) R(t,s) = 0 
\nonumber \\
X_1 R &=& \left( -t^2\partial_t -s^2\partial_s -\frac{2}{z}\left(x+\xi\right)t
-\frac{2}{z}\left(\wit{x}+\wit{\xi}\right)s\right) R(t,s) = 0
\EEA
where $\wit{x}$ and $\wit{\xi}$ refer to the response field $\wit{\phi}$. 
To solve these, change variables into $u=t-s$ and $v=t/s$. Then, with
$R(t,s)=\bar{R}(u,v)$
\BEA
\left( u\partial_u +\frac{x+\wit{x}}{z}\right) \bar{R}(u,v) &=& 0 
\nonumber \\
u \left(v\partial_v +\frac{v}{v-1}\frac{x-\wit{x}+2\xi}{z}
+\frac{1}{v-1}\frac{\wit{x}-x+2\wit{\xi}}{z}\right) \bar{R}(u,v) &=& 0
\EEA
The solution to these is found in factorized form $\bar{R}(u,v)=f(u)g(v)$ and 
we find, after having returned to the variables $t$ and $s$
\BEQ
R(t,s) = r_0 s^{-1-a} \left( \frac{t}{s}-1\right)^{-1-a'} 
\left(\frac{t}{s}\right)^{1+a'-\lambda_R/z}
\EEQ
where
\BEA
1+a &=& \frac{x+\wit{x}}{z} \nonumber \\
1+a' &=& \frac{x+\wit{x}+2\xi+2\wit{\xi}}{z} \\
\lambda_R &=& 2\left( x+\xi\right) 
\nonumber
\EEA
which is the form (\ref{gl:Rf}) stated in the text. 
We finally observe that if $\xi+\wit{\xi}=0$, we recover indeed $a=a'$. 

%%%%%%%%%%%%%%%%%%%%%%%%%%%%%%%%%%%%%%%%%%%%%%%%%%%%%%%%%%%%%%%%%%%%%%%%%%%%%%%%
\appsection{B}{On the Ohta-Jasnow-Kawasaki approximation}
%%%%%%%%%%%%%%%%%%%%%%%%%%%%%%%%%%%%%%%%%%%%%%%%%%%%%%%%%%%%%%%%%%%%%%%%%%%%%%%%

We briefly present the derivation of the scaling of the response function 
$R(t,s)$ in a simple analytically tractable scheme which is close in spirit
to the Ohta-Jasnow-Kawasaki (OJK) approximation. Following the ideas 
of Berthier, Barrat and Kurchan \cite{Bert99} and of Mazenko \cite{Maze04} 
which in turn are based on a calculation by Bray \cite{Bray97}, 
one considers the integrated response in the zero-field cooled
protocol. When perturbing the system by a spatially random field $h$ of
magnitude $h_0$, the zero-field-cooled (ZFC) susceptibility 
reads \cite{Bert99,Bray97} 
\BEQ
\chi(t,s) = \frac{\langle h(\vec{r})\phi(t,\vec{r})\rangle}{h_0^2} 
\simeq \sqrt{\frac{2}{\pi}\,} \frac{\langle
h(\vec{r})m(t,\vec{r})\rangle}{h_0^2 \sqrt{ \langle m^2\rangle}} 
\EEQ
where $\phi$ is the order parameter and  it is assumed that for late times, one
can approximate $\phi\sim \mbox{\rm sign }(m)$. At zero temperature, the
auxiliary field $m$ should satisfy the equation of motion
\BEQ
\frac{\partial m}{\partial t} = \nabla^2 m - n_a n_b \nabla_a \nabla_b m
+ h |\nabla m|.
\EEQ
In the spirit of the OJK-approximation, one makes \cite{Bray97,Bray94} 
the simplifications $n_a n_b \to \delta_{ab}/d$ (circular average) and 
$|\nabla m| \to \langle (\nabla m)^2\rangle^{1/2}$. The equation of motion
then becomes \cite{Bert99,Bray97} 
$\partial_t m = D \nabla^2 m + h\langle (\nabla m)^2\rangle^{1/2}$ with
$D=(d-1)/d$ and assuming the fields $m$ and $h$ to be gaussian, it is found 
that \cite[eqns.~(19,20)]{Bert99}
\BEQ \label{Bgl:chi1}
\chi(t,s) = \int_s^t \!\D u\: 
\frac{(D t)^{d/4}}{(D u)^{(d+2)/4}}
\int_{\mathbb{R}^d} \!\D\vec{k}\: e^{-k^2 D(t-u)}
\EEQ
which is the starting point of our analysis. 

Performing the integration in $\vec{k}$-space, one obtains \cite{Maze04}
\BEQ \label{Bgl:chiR}
\chi(t,s) = \mbox{\rm cste.} \int_s^t \!\D u\: u^{-(d+1)/2} 
\left(\frac{t}{u}\right)^{d/4} \left( \frac{t}{u} -1 \right)^{-d/2}
\stackrel{!}{=} \int_s^t \!\D u\: R(t,u).
\EEQ
This integral becomes singular near the upper limit $u\approx t$ and we shall
reconsider this below. Before we shall do this, we read off the autoresponse
function
\BEQ \label{Bgl:R}
R(t,u) = u^{-(d+1)/2} f_{{\rm OJK}}(t/u) \;\; , \;\; 
f_{{\rm OJK}}(y) = f_0\, y^{d/4} (y-1)^{-d/2}
\EEQ
{}from which we easily recover, using eq.~(\ref{gl:Rf}), the values of the 
exponents $a$, $a'$ and $\lambda_R$ in the OJK-approximation (where $z=2$) 
as stated in eq.~(\ref{gl:expo}). Eq.~(\ref{Bgl:R}) is also recovered
in the gaussian theory of phase-ordering \cite{Maze03}. 

A different conclusion was reached by Berthier et {\em al.} \cite{Bert99}
albeit starting from the same eq.~(\ref{Bgl:chi1}). They quote
$\chi(t,s) \sim s^{-1/2} F(t/s)$ for $d>2$ which would mean $a=1/2$, 
provided of course that the na\"{\i}ve scaling law 
$\chi(t,s)=s^{-a}f_{\chi}(t/s)$ could 
be used. We must therefore reconsider the singularity in eq.~(\ref{Bgl:chiR}) 
for $\chi(t,s)$. Following \cite{Bert99,Maze04}, 
one introduces a cut-off parameter $\Lambda^2$ (which should be sent to 
zero at the end) and writes instead of (\ref{Bgl:chi1})
\BEQ \label{Bgl:chi2}
\chi(t,s) = \int_s^t \!\D u\: 
\frac{(D t)^{d/4}}{(D u)^{(d+2)/4}}
\int_{\mathbb{R}^d} \!\D\vec{k}\: e^{-k^2 D(t-u+\Lambda^2)}.
\EEQ
Performing first the integral over $\vec{k}$ and changing variables, 
this becomes, up to a normalization constant 
\BEQ
\chi(t,s) \sim t^{(1-d)/2} \int_{s/t}^1 \!\D v\: v^{-(d+2)/4} \left( 1 -v
+\Lambda^2/t\right)^{-d/2}
\EEQ
and we must analyze the contribution of the integrand near $v\approx 1$. We 
decompose the domain of integration $\int_{s/t}^1 = \int_{s/t}^{1-\eps}
+ \int_{1-\eps}^1$. In the first term, we let $\Lambda^2\to 0$, and
$v\simeq 1$ in the first factor of the second term. Then, for $d>2$
\BEA
\chi(t,s) &\simeq& t^{-(d-1)/2} \int_{s/t}^{1-\eps} \!\D v\: 
v^{-(d+2)/4} (1-v)^{-d/2} 
\nonumber \\
& & + t^{-(d-1)/2} \int_{1-\eps}^1 \!\D v \:
\left( 1 - v + \Lambda^2/t\right)^{-d/2} 
\nonumber \\
&=& t^{-a} F(s/t) + \mathfrak{c}_1 t^{-1/2} \left( \Lambda^2\right)^{1-d/2}
+ \mathfrak{c}_2 t^{-a} \left( \eps +\Lambda^2/t\right)^{1-d/2}
\nonumber \\
&=& s^{-1/2} \cdot \chi_{\infty} \left(t/s\right)^{-1/2} 
+ s^{-a} f_{\chi}(t/s)
\EEA
where the value $a=(d-1)/2$ was used. Here $\mathfrak{c}_{1,2}$ and
$\chi_{\infty}$ are constants and $F$ and $f_{\chi}$ are scaling functions. 
In this way, not only we recover the leading term already found in 
\cite{Bert99} but we also see that the contribution coming form formally
integrating the scaling behaviour of $R(t,s)$ merely gives rise to a
sub-leading correction. A similar argument can be applied to the case
$d=2$ and produces logarithmic corrections. 

Consequently, the mere observation of a scaling law for $\chi(t,s)$ in the
scaling regime $t,s\to \infty$ with $y=t/s$ fixed is not enough to reliably
extract the exponent $a$. Indeed, the term expected from na\"{\i}ve scaling 
$\chi(t,s)\sim s^{-a}$ merely arises as a short-time correction to 
the leading long-time behaviour
of $\chi(t,s)$. It has already been pointed out in the
context of the Ising model quenched to $T<T_c$ that a similar dominant term
not simply related to the autoresponse exponent $a$ occurs in the 
ZFC-susceptibility for $d\leq 3$ \cite{Henk03e}. The straightforward use of 
$\chi(t,s)$ may hence lead to erroneous values of the exponent $a$ and it is
safer to avoid using $\chi(t,s)$ altogether.

\newpage

%%++++++++++++++++++++++++++++++++++++++++++++++++++++++++++++++++++++++++++++++

%%%%%%%%%%%%%%%%%%%%%%%%%%%%%%%%%%%%%%%%%%%%%%%%%%%%%%%%%%%%%%%%%%%%%%%%%%%%%%%%

\end{document}